\documentclass[mathleft
]{an}
\usepackage{graphicx}

\usepackage{multicol}
\usepackage{wrapfig}
\usepackage{natbib}

\usepackage{graphics}

\usepackage{times}

\newcommand{\be}{\begin{equation}}
\newcommand{\ee}{\end{equation}}

\newcommand{\ba}{\begin{eqnarray}}
\newcommand{\ea}{\end{eqnarray}}

\newcommand{\Alfven}{ Alfv\'{e}n }

\newcommand\eg{\textit{e.g.,\ }}

\newcommand{\Bf}{{magnetic field}}
\newcommand{\Bfs}{{magnetic fields}}
\newcommand{\Ef}{{electric  field}}
\newcommand{\Efs}{{electric fields}}

\newcommand{\ms}{magnetosphere}
\newcommand{\mss}{magnetospheres}
\newcommand{\Fermi}{{\it Fermi}}

\begin{document}

% The following seven commands are intended for editorial usage and should be ignored by
% the author(s).
%\Pagespan{789}{}% Document's page range. 
% If second parameter is left empty, the last page is computed automatically.
%\Yearpublication{2006}%
%\Yearsubmission{2005}%
%\Month{11}%   
%\Volume{999}%  
%\Issue{88}% 
% \DOI{This.is/not.aDOI}% 

\title{I. Inverse Compton origin of pulsar $\gamma$-ray emission.\\
II.  Reconnection model of Crab flares.}

\author{Maxim Lyutikov\inst{1}\fnmsep\thanks{Corresponding author:
  \email{lyutikov@purdue.edu}\newline}
%Example 
%for footnote, note the usage of the \texttt{fnmsep}
%command as separator between institute number and footnote mark} 
}
\titlerunning{Pulsar and Nebula  gamma-ray emission}
\authorrunning{M. Lyutikov}
\institute{Department of Physics, Purdue University, 
 525 Northwestern Avenue,
West Lafayette, IN
47907-2036}

%\received{30 May 2005}
%\accepted{11 Nov 2005}
%\publonline{later}

%\keywords{Editorial notes -- instruction for authors}

\abstract{I. There is growing evidence that pulsars' high energy  emission is generated via Inverse Compton  mechanism. 
\\
II.  The  particles producing  Crab flares, and possibly most of the Crab Nebula's high energy emission, are accelerated  via reconnection events, and not  at shock via Fermi mechanisms.
}

\maketitle

\section{PULSAR   $\gamma$-RAY EMISSION}

\begin{figure}[h!]
\includegraphics[width=.99\linewidth]{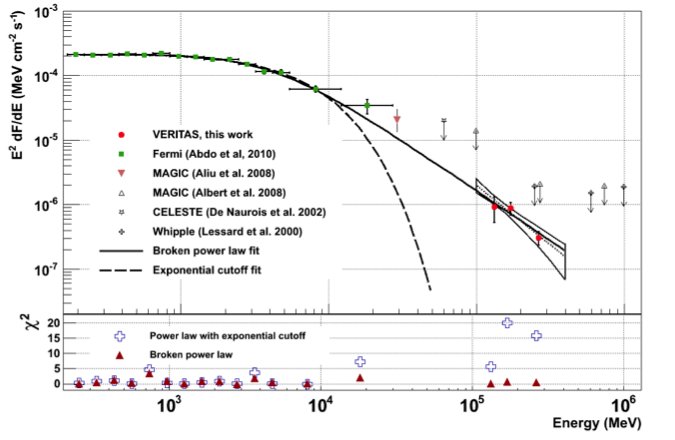}\\
\includegraphics[width=0.99\linewidth]{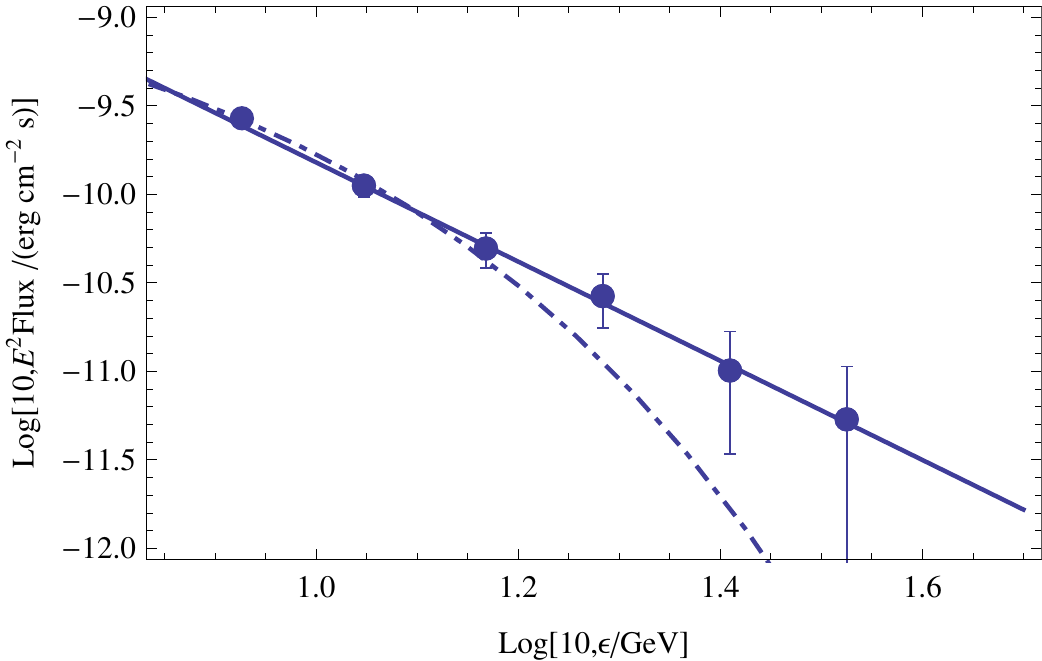}
\caption{Above. Spectral energy distribution  of the Crab pulsar \cite{VERITASPSRDetection}. Below.
 Fits to the high energy tail of the  Geminga spectrum: power law (solid line) and exponential cut-off (dashed line) \cite{2012ApJ...757...88L}. In both cases the high energy part of the  spectrum is power-law, inconsistent with curvature mechanism.}
\label{spectrum-Crab}
\end{figure}

Conventionally,   $\gamma$-ray from pulsars were attributed to incoherent curvature radiation. 
The detection of the Crab pulsar by VERITAS  \citep{VERITASPSRDetection}, Fig. \ref{spectrum-Crab}, and the implied importance of inverse Compton (IC) scattering for the production of $\gamma$-ray photons  \citep{2012ApJ...754...33L}  signifies a paradigm shift  in the study of pulsar high-energy emission.   Most models of the underlying physical processes have to be re-evaluated (the geometrical models  remain valid.) The importance of the IC scattering makes distinct predictions as to the observed properties of pulsar high-energy emission. 

 One can derive  a very general upper limit, independent of the particular details of the acceleration mechanism,  on the possible energy  of curvature photons \citep[following  a similar approach applicable to synchrotron emission][]{1996ApJ...457..253D,2010MNRAS.405.1809L}. 
Assuming  that the  radius of curvature $R_c $ of the  \Bf\ lines  is a fraction  $\xi$ of the light cylinder radius $R_c = \xi R_{LC}$ and 
balancing acceleration by \Ef\ of the fraction of the \Bf, $E=\eta B$ and radiative losses
 one can find {\it the maximum energy of curvature emission within the Crab pulsar \ms}
$
\epsilon_{br} 
%(3 \pi)^{7/4}  { \hbar \over ( c e) ^{3/4} } \eta^{3/4} \sqrt{\xi} \,  {  B_{NS}^{3/4} R_{NS}^{9/4}\over P^{7/4}}
 =
150\, {\rm GeV} \, \eta^{3/4} \sqrt{\xi} .
% = 5\, {\rm GeV} \, \eta_{-2} ^{3/4} \sqrt{\xi} 
% \label{1}
$
 The  possibility  that the emission above the break energy is produced  as a tail to the curvature emission  is excluded by the fact that 
 the spectrum of Crab pulsar reported by  VERITAS {\it does not show the exponential cut-off} indicative of  the radiation reaction-limited  curvature emission.
 
 Reanalyzing  the Fermi
  spectra of the Geminga   pulsar  above the break,   \cite{2012ApJ...757...88L}  found that it is well approximated by  a simple power law  without the exponential cut-off, making  Geminga's spectrum  similar   to that of  Crab, Fig. \ref{spectrum-Crab}.   Vela's broadband $\gamma$-ray spectrum is equally well fit with both the exponential cut-off and the double power law shapes.  In  the broadband double power-law fits,  for a typical Fermi spectrum of a bright $\gamma$-ray pulsar,  most of the errors accumulate due to the  {\it arbitrary} parametrization of the spectral roll-off. Thus, based on the data from the first pulsar catalogue, {\it  out of three brightest $\gamma$-ray pulsars, the spectra of  two (Crab and Geminga) are inconsistent with the exponential cut-off, while the third (Vela) is consistent with double power-law.}

 \cite{2012ApJ...754...33L} 
 demonstrated that  Inverse Compton scattering by the secondary particles  in  a Klein-Nishina regime is broadly consistent with the overall energetics and emitted spectra.  Klein-Nishina reduction in the scattering cross-section (and the corresponding  reduction
of the electron energy loss rate)  allows the primary  leptons to be accelerated to very high energies with hard spectra. The secondary plasma is less energetic but more dense and has approximately the same energy content as the primary beam, producing Synchrotron-Self-Compton radiation. The  synchrotron component has a broad peak in the UV-$X$-ray range, while the IC component extends to hundreds of {\rm GeV}. Below $\sim 1$ {\rm GeV} the  curvature emission from the primary beam can contribute a substantial flux fraction. The IC emission from the primary bean extends well into the {\rm TeV} regime but can hardly   be detected by the ground Cherenkov telescopes.

\begin{figure}
\includegraphics[width=.99\linewidth]{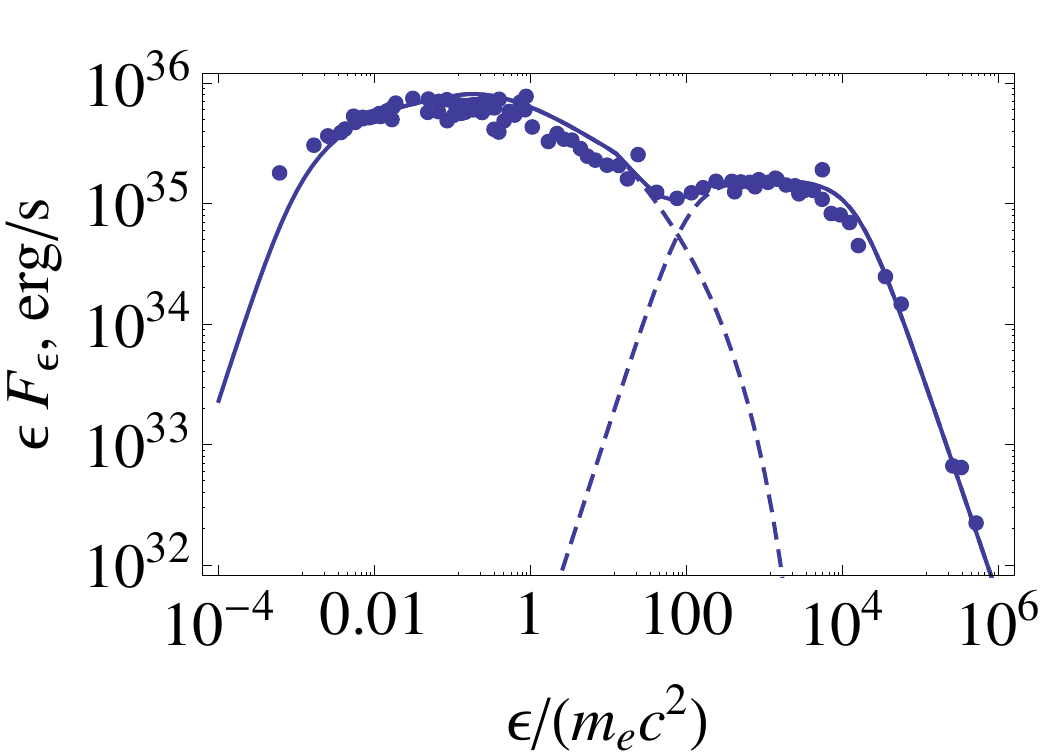}
\caption{ The broadband spectrum of the Crab approximated with the CSC model  \citep{2013MNRAS.431.2580L}. The IC bump in the KN regime provides a direct measurement of the bulk  particle distribution, while the high energy part of  cyclotron bump constrains the very high energy tail of the particle distribution.
 {\bf This is a fit over nearly ten decades in energy, using only a handful of parameters.}}
\label{CrabPulsarFit}
%\end{figure}
\end{figure} 
The model   determines the time and phase averaged outcome of the complicated pair production and acceleration processes. Typical Lorentz factor turn put to be $\gamma \sim $  thousands,  which is very reasonable \cite{HardingMuslimov98}. The only difference with the conventional view is high plasma multiplicity, $\lambda \sim  10^6 - 10^7$, but, recall,  that Crab nebula needs $\lambda \sim   10^6$ on average. The procedure to determine the particle spectrum from the observed photon spectrum can be fully numerical (without, \eg\ assumption of a power law), with some non-analytical result (in KN regime particle spectrum is easily related to the photon spectrum).

{Our  results indicate  \citep{2013MNRAS.431.2580L} that  the broadband  spectrum of  Crab pulsar, from UV to very high energy gamma-rays}  - nearly ten decades in energy - can be reproduced  within the framework of  the cyclotron-self-Compton model. Emission is produced  by  two counter-streaming beams within the outer gaps, at distances above  $\sim$ 20 NS radii. The outward moving beam  produces UV-$X$-ray photons via Doppler-booster cyclotron emission, and GeV photons  by Compton scattering the cyclotron photons produced by the inward going beam. The scattering occurs in the deep Klein-Nishina regime, whereby the  IC component provides a direct measurement of particle distribution within the magnetosphere.   The required plasma multiplicity is high, $\sim 10^6-10^7$, but is consistent with the average particle flux injected into the pulsar wind nebula, Figs.  \ref{CrabPulsarFit}. 

\begin{figure}
\vskip -0.15 truein
\centering
\includegraphics[width=.99\linewidth]{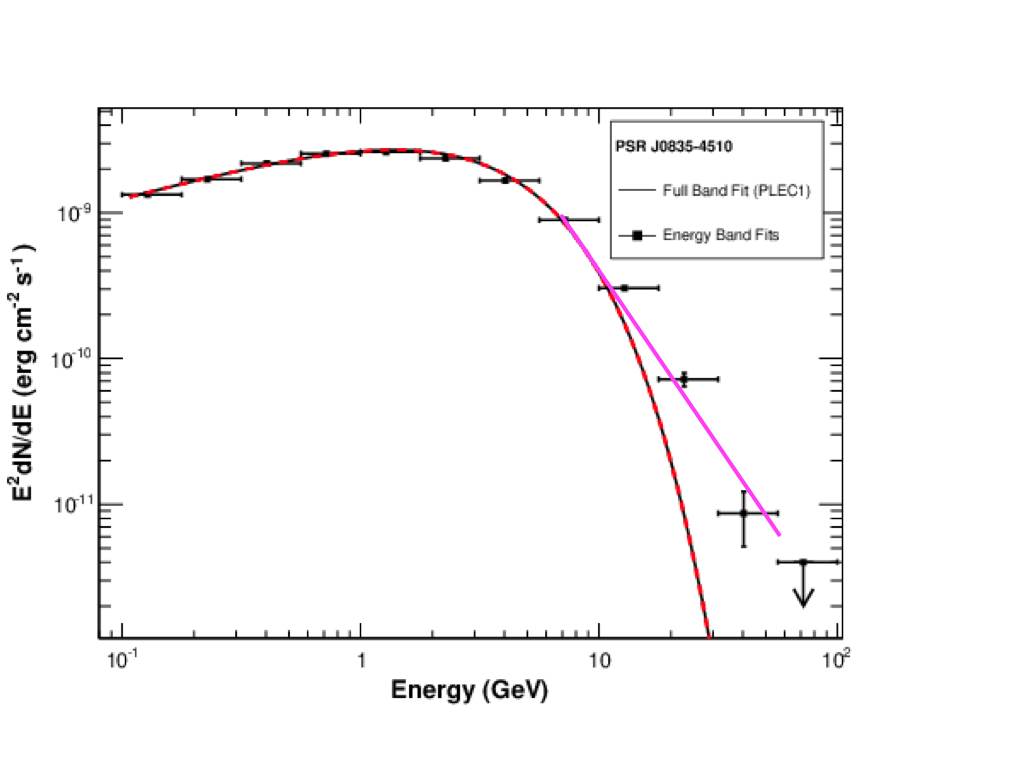} \\  \hskip -.5truein \includegraphics[width=.89\linewidth]{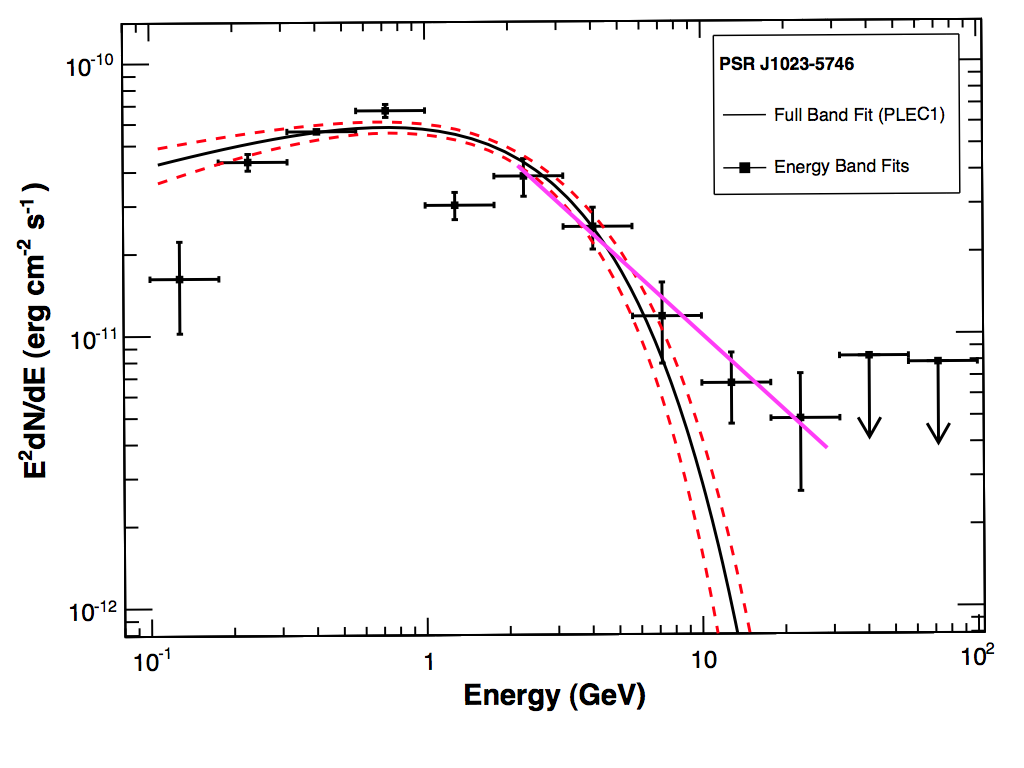} \\
\includegraphics[width=.89\linewidth]{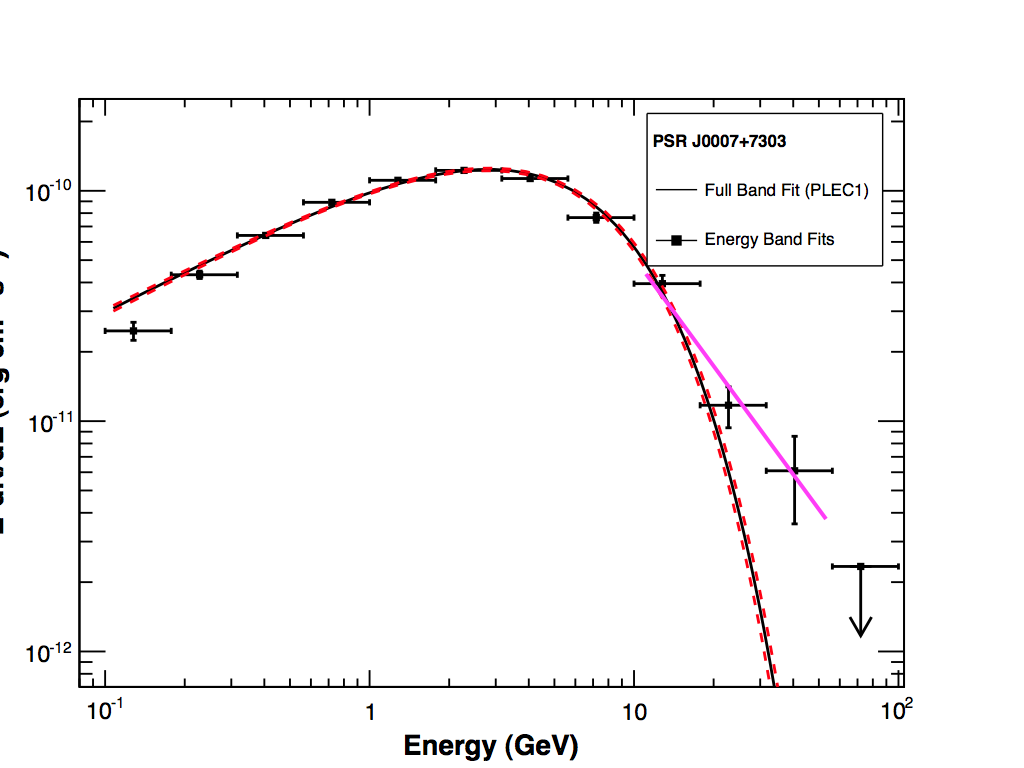}  \\ \hskip -.5truein   \includegraphics[width=.95\linewidth]{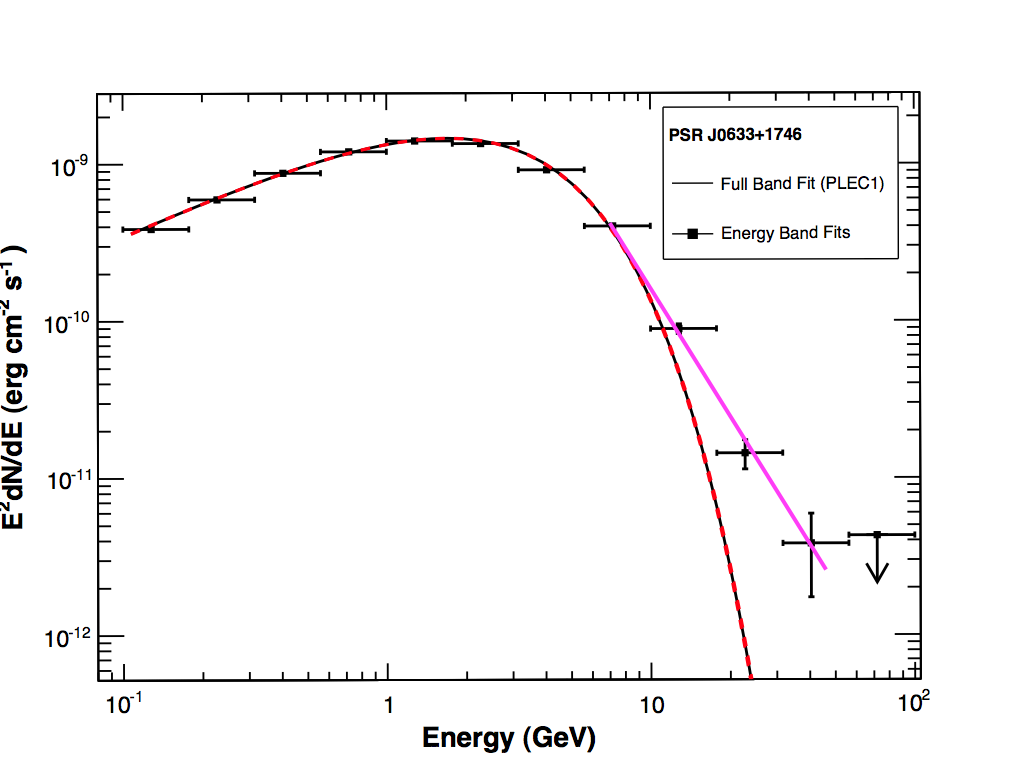}
\caption{Example of pulsar spectra from the second Fermi-LAT catalogue indicative of high energy  power-law tails. Red lines: power law plus exponential fits \citep{2013arXiv1305.4385T}, magenta lines: power law fits for high energy tails.}
%\vskip -.5truein
\label{Fits}
\end{figure}

There are indications in the second Fermi-LAT catalogue that at high energies pulsar spectra and described better by the power law, and the exponential cut-off spectral shape, \ref{Fits}. 

\begin{figure}
\begin{center}
\includegraphics[width=.99\linewidth]{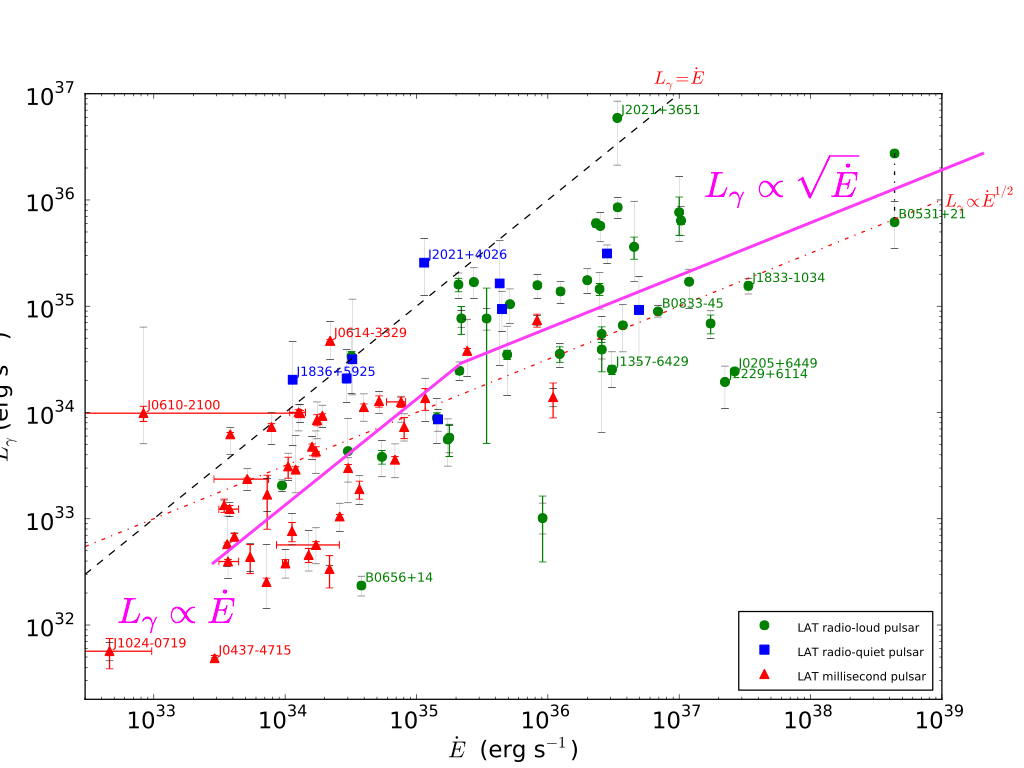}
%\vskip -1.2 truein
\end{center}
\caption{The proposed scaling of the  $L_\gamma (\dot{E})  $ correlation is linear in the radiation reaction-limited regime (small $ \dot{E}  $) and square root in the potential-limited regime (large $ \dot{E}  $).
}
\label{LumiVSEdot}
\end{figure}

{\bf Scaling of $\gamma$-ray luminosity with spin-down power.}
Commonly, the $\gamma$-ray luminosity is thought to scale with the available potential, which is  the square root of  the  spin-down luminosity, i.e. $\propto \sqrt{ \dot{E}}  $.  Lyutikov \& Otte (2010) demonstrated that 
 {\it  in the radiation reaction-limited regime, the very high energy luminosity is proportional to the  spin-down luminosity}:
 \be
 L_\gamma =   \eta \eta_G { B_{NS}^2 R_{NS}^6 \Omega^4 \over 2 \pi c^3} =  \eta \eta_G \dot{E}
 \label{11}
\ee
 where $\eta_G$ is the volume fraction occupied by the outer gaps, in terms of the light cylinder volume. 
   Recall, that the square-root  scaling comes from the assumption (initially appropriately used in the polar cap models of pulsar $\gamma$-ray emission) of a potential-limited energy of accelerated particles,   that most of this energy is radiated away, and that  the  particle flux is of the order of the Goldreich-Julian density, 
$\dot{N} \propto n_{GJ} r_{PC}^2 \propto \sqrt{\dot{E}}$ \citep[see \eg][]{2000ApJ...532.1150Z}.   In contrast, 
in a radiation-limited regime, and assuming that the 
emitting volume is proportional to the volume within the  light cylinder radius, one obtains    $L_\gamma \propto \dot{E}  $ (Lyutikov \& Otte, 2011), Fig. \ref{LumiVSEdot}.  
%\begin{figure}
%\end{figure}
The estimate (\ref{11}) of the $\gamma$-ray luminosity does not depend on the particular radiation mechanism that limits the particle energy; it will be the same estimate if, \eg, curvature losses or  IC scattering  in the Thompson regime provide the dominant drag on the particle. On the other hand, it is not really applicable if the sole radiative loss mechanism is the   IC scattering  in KN regime. In this case  the radiative losses are independent of the particle energy, so that 
the  acceleration by the \Ef\  cannot be balanced in  a general case  by the radiative losses. 

{\bf Two photon pair production in  outer gaps}.
The models of the pulsar \mss\ are mostly based on the  \cite{1975ApJ...196...51R} model of pair production near the polar caps. It was then expected that polar cap regions are intense sources of high energy emission \citep{1982ApJ...252..337D}. These expectations are not supported by few years of \Fermi\ data,  no gamma-ray emission from pulsar polar caps have been observed. On the other hand,  the Crab nebula $X$-ray emission requires a huge flux of particles, with the {\it average} over the open field lines multiplicity of $<\lambda> \sim 10^6$. Thus, there is a  contradiction between polar cap pair production models and observations. The importance of  Compton scattering in the Klein-Nishina regime  implies the importance of  pair production in the outer gaps: the corresponding cross-sections are very similar. We will investigate  whether a self-consistent model of particle acceleration and pair production in the gaps leads to a stationary particle distribution (this is a highly non-linear electrodynamic problem).

\begin{figure}
%\begin{flushleft}
\begin{center}
% --------- NEW lines -----------
\includegraphics[width=.8\linewidth]{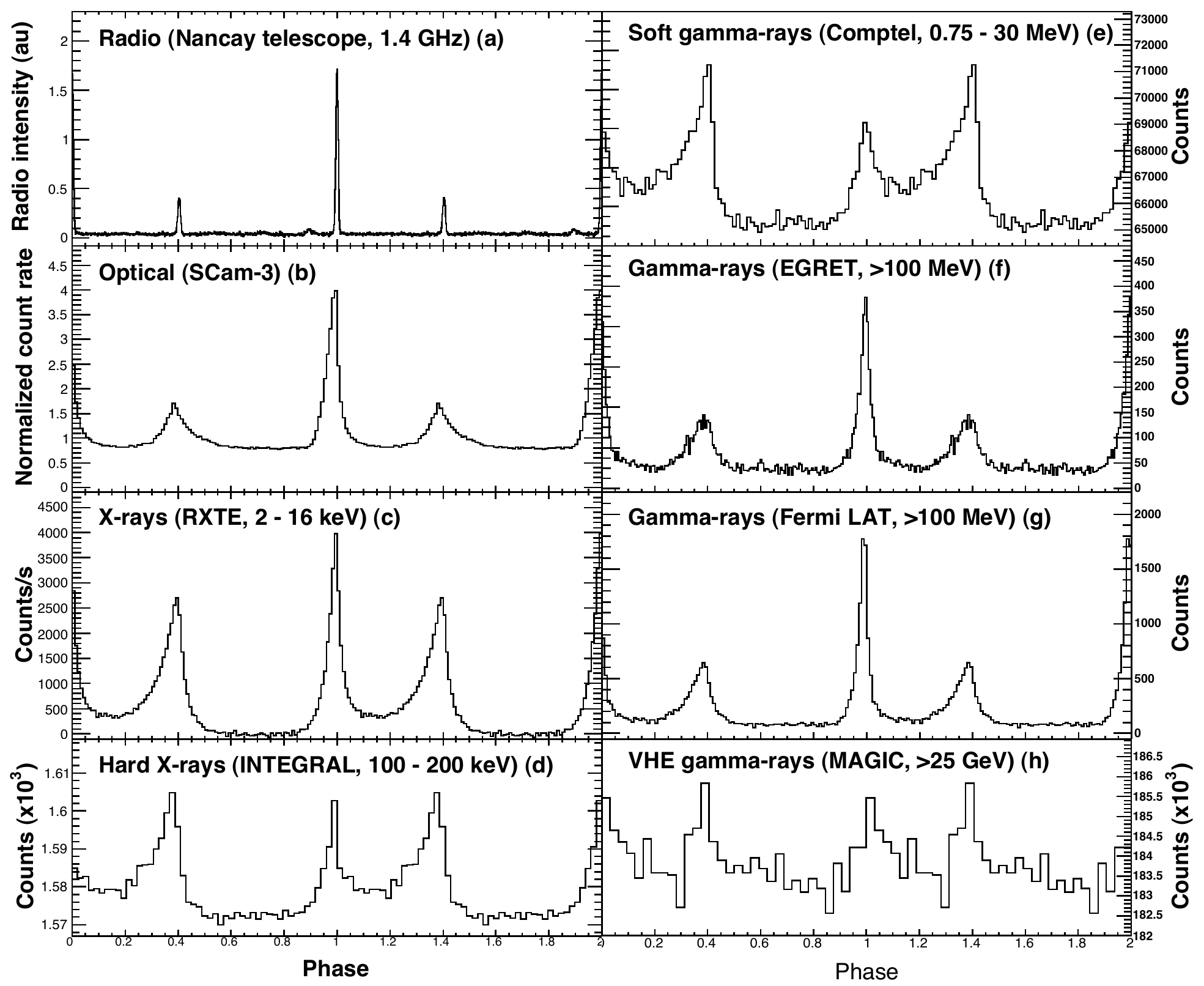}
\label{Crab-profile-Fermi}
\caption{
The average profile of the Crab pulsar from radio to $\gamma$-rays \citep[from][]{2010ApJ...708.1254A}. The ratio of the interpulse to main pulse is repeated in the two spectral band:  low in optical,   increases  to soft $\gamma$-rays, becomes low again in the MeV range and starts to increase again at GeV energies. In the IC model the two radiative components are expected to mirror each other.
}
\end{center}
\end{figure}

{\bf  Expected  $X$-ray-$\gamma$-ray correlation}.
Since the IC power is related to the power of the seed photons, we might expect some $\gamma$-ray - soft $X$-ray correlation. Though it is the soft UV photons that are scattered to {\rm GeV} energies, and, formally, one expects UV-{\rm GeV} correlation,   $X$-rays and  UV photons form a continuous spectral distribution and so one also expects $X$-ray - {\rm GeV} correlation. One such correlation is well known: the ratio of MP to IP changes with energy and these changes are repeated in the two spectral bumps, Fig. \ref{Crab-profile-Fermi}

{\bf  Relating radio to high energy  emission.} The present model of pulsar high energy  emission requires that the emitting particles have a finite pitch angle. 
A possible excitation mechanism is  related to the generation of the pulsar radio emission at the anomalous cyclotron resonance \citep{lbm99}. In this model
the pulsar radio emission is  produced directly
 by   maser-type
plasma instabilities operating at the anomalous cyclotron-Cherenkov resonance
$\, \omega-\, k_{\parallel} v_{\parallel} +\, |\omega_B|/\,  \gamma_{res}=0$ (note the sign in front of the cyclotron term).
At the anomalous resonance a particle emitting a wave undergoes a transition {\it  up} (!)  in Landau levels. Thus,  initially one-dimensional distribution develops a finite pitch angle. As a result, resonant particles start emitting  cyclotron photons at the normal cyclotron resonance.

Cyclotron motion of particles  in the pulsar magnetosphere may be excited due to  coherent  emission of radio waves by  streaming particles at the anomalous cyclotron resonance. Thus, a whole range of  Crab non-thermal emission, from coherent radio waves to very high energy $\gamma$-rays - nearly eighteen  decades in energy - may be a manifestation of inter-dependent  radiation processes.

\section{ THE CRAB NEBULA GAMMA-RAY FLARES}

\begin{figure}
\includegraphics[width=.99\linewidth]{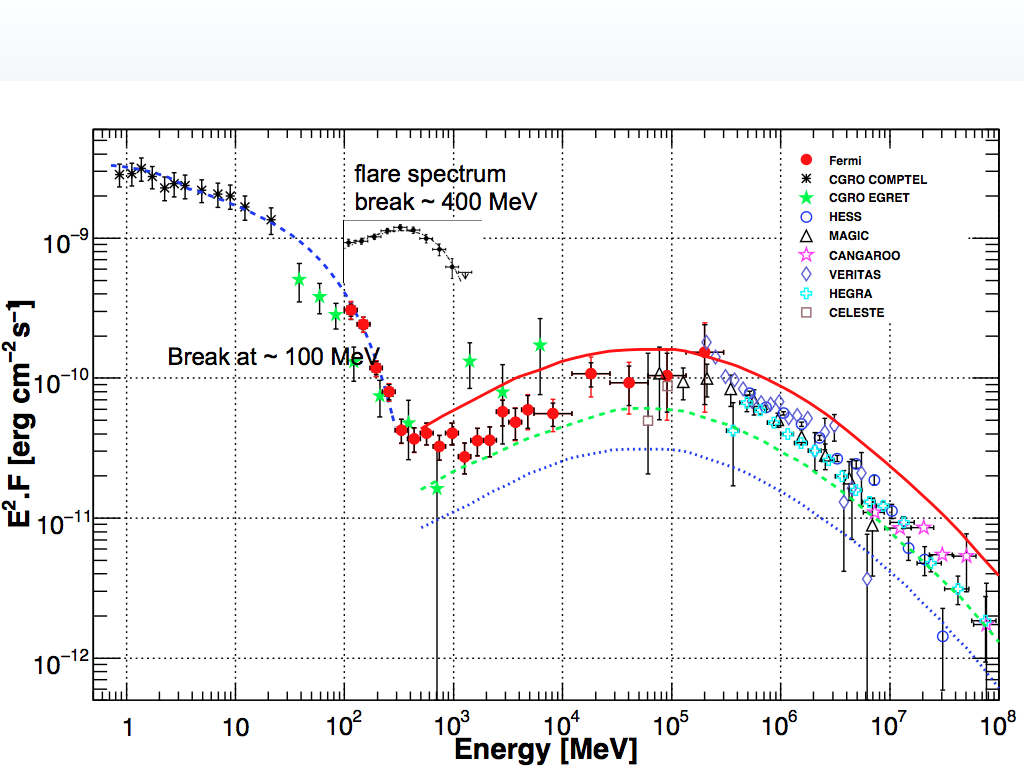}
\caption{\small Crab spectrum with superimposed flare spectrum.}
\label{SpectrumCrabAndFlare}
%\end{figure}
\end{figure}

The detection of  flares from the Crab Nebula by AGILE and Fermi satellites  \citep{2011Sci...331..736T,2012ApJ...749...26B}, Fig. \ref{SpectrumCrabAndFlare},   is one of the most astounding discoveries in high energy astrophysics. 
The unusually short durations, high luminosities, and high photon energies of the Crab Nebula gamma-ray flares require reconsideration of our basic assumptions about the physics processes responsible for acceleration of (some) high-energy emitting particles in the Crab Nebula, and, possibly in other high-energy  astrophysical sources. During flares, the Crab Nebula gamma-ray flux above 100 MeV exceeded its average value by a factor of several or higher \citep{2011Sci...331..736T,2011Sci...331..739A}, while in other energy bands nothing unusual was observed.

We associate the duration of the flare with stochastically changing local properties of plasma within the nebula. We envision flares originating from a highly localized emission region, so that the flare observables determine the intrinsic properties of the emission region, and not necessarily the overall bulk properties of the flow. The natural flaring mechanism in this category is random relativistic magnetic reconnection
 events \citep{2012MNRAS.426.1374C}.

{\bf  Acceleration in PWNe: shocks and/or reconnection?}
 The discovery of flares 
  from  
Crab Nebula challenges our understanding and particle acceleration in the PWNe. On the one hand,   
   the low magnetization numerical models of Crab Nebula  \citep{KomissarovLyubarsky,DelZanna04} are able to reproduce the morphological details of the Crab Nebula down to intricate details. The implied acceleration mechanism is then the  acceleration at the termination shock  \citep{reesgunn,kc84}, presumably via the stochastic Fermi-type process. This is currently the dominant paradigm.

 On the other hand, even  before the discovery of Crab flares, \cite{2010MNRAS.405.1809L} argued that the
 observed cutoff in 
synchrotron spectrum of the Crab Nebula at  $\sim 100$ MeV in the persistent Crab Nebula emission  is probably too high to
be consistent with the stochastic shock acceleration. Indeed, balancing electrostatic acceleration in a regular electric field with 
synchrotron energy losses 
 yields the synchrotron photon energy 
$
\epsilon_{\rm max} \sim \eta   \hbar  { m c^3 \over e^2} \approx 100 \mbox{ MeV},
\label{emax}
$
where $\eta$ is the ratio of electric to magnetic 
 field strengths. Since high conductivity of astrophysical plasma ensures that in most 
circumstances $\eta<1$, the observed value of the cutoff is right at the very limit.  
During the gamma-ray flares the cutoff energy approached even higher value of 
$\sim 400 MeV$, suggesting Doppler boosting or a different 
acceleration mechanism. In addition, recent optical observations show no significant 
variability of the brightest features associated with the termination shock 
\citep{2012arXiv1211.3997W}, apparently ruling out the shock acceleration as a mechanism 
behind the flares.

\cite{2013MNRAS.428.2459K,2012MNRAS.427.1497L}  pointed out that while particle acceleration 
may be efficient at the termination shock of the equatorial striped section of pulsar 
wind, it is unlikely to operate in the polar section which is free from stripes. 
 Moreover, the magnetic dissipation is needed to reconcile the observed 
sub-equipartition magnetic field of the Crab Nebula with its theoretical models.    
This conclusion is supported by the recent numerical simulations \cite{2013MNRAS.431L..48P}. 
Thus, low magnetization models reproduce the Crab PWN morphology, but the implied acceleration mechanism, the stochastic shock acceleration, fails to reproduce the spectrum. 

\cite{2012MNRAS.426.1374C} argued  that  the flares can be due to a highly localized emission region, or blob, so that the flare observables determine the intrinsic properties of the emission region. 
Magnetic reconnection provides a natural explanation of the implied relativistic motion discussed above, the intrinsic short timescales, and the flares' intermittency. Reconnection is a process in which the magnetic energy of a localized region, a current sheet, is suddenly converted to random particle energy, and bulk relativistic motion.  \cite{2012MNRAS.426.1374C} demonstrated that in the radiation reaction limit  the electron distribution function may display a power-law with an excess, or pile-up, of particles near the synchrotron limit, in which case the emitting particles will display a SED that is close to the single-particle synchrotron SED.  The observed spectrum of the flare is consistent with a mono-energetic pile-up \cite{2012MNRAS.426.1374C}.
\begin{figure}
\centering
\includegraphics[width=.99\linewidth]{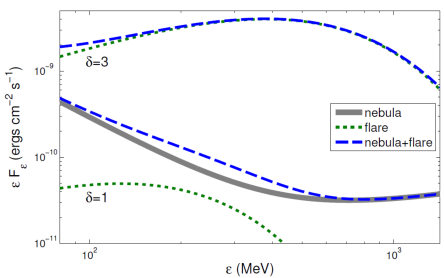}
\caption{ Doppler boosted single-particle synchrotron SEDs (dotted lines) for $\delta=1$ and $\delta= 3$.  The average nebula SED (thick line) is summed with the flare SEDs to produce a combined SED shown by the dashed line.}
\label{spectrum} 
\end{figure}

Even a mild Doppler beaming can resolve the problem with the synchrotron upper limit. In Fig.  \ref{spectrum} we have plotted  the same  small (total number of emitting electron $\sim 10^{38}$, produced by  a pulsar within a second) intrinsic single-particle synchrotron SED with two different Doppler factors, $\delta=1$ and $\delta= 3$.  The different Doppler factors affect the intrinsic SED's photon energies, $\epsilon=\delta \epsilon'$, and normalization, $\epsilon F_{\epsilon}(\epsilon)=\delta^4 \epsilon'F'_{\epsilon'}(\epsilon')$ \citep{Lind:1985}.  

The non-detection of significant flaring by X-ray telescopes places a further constraint on the flare SEDs.  If the flare SED comes from a mono-energetic electron distribution, then in the \textit{Chandra} and \textit{XMM-Newton} energy bands, the SED goes as $\epsilon  F_{\epsilon}\propto \epsilon^{4/3}$.  For the single-particle synchrotron SED, the X-ray variations are generally non-detectable.   

\begin{figure}
\centering
\includegraphics[width=.99\linewidth]{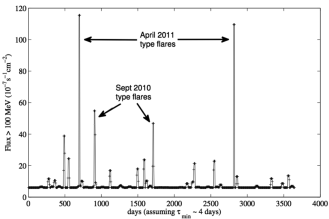}
\caption{Monte-Carlo simulations of Crab Nebula flux variations with the minijet model.  The  minijet flux averaged over different flares \textit{is dominated by rare bright flares}.}
\label{Flares-MC} 
\end{figure}

{\bf The  minijet  model.}  We identify  the flaring as  reconnecting sites  in the nebula plasma. The reconnection outflow speeds are then relativistic \citep{LyutikovUzdensky} and behave as ``minijets," a model that has been used to overcome the gamma-ray opacity problem in the context of gamma-ray bursts and active galactic nuclei \citep{LyutikovRem}.  We have constructed a statistical model of Crab Nebula high energy variability by assuming magnetic reconnection sites throughout the nebula are activated randomly, and once activated, display emission characteristics controlled by the reconnection outflow Doppler factor \citep{2012MNRAS.426.1374C}, Fig. \ref{Flares-MC}.  In our model, GeV flares are observed when a magnetically dominated site in the Crab nebula undergoes magnetic reconnection and launches a relativistic outflow that is, by chance, aligned with the line of sight.  The observed flares' unusually short durations and high luminosities suggest the emitting plasma is indeed moving toward Earth at relativistic speeds.
Assuming shot-noise type generation of minijets, we can calculate observable quantities such as minijet timescale and flux distributions \citep{2012MNRAS.426.1374C}. Importantly, due to high powers dependence of the flux on Doppler beaming,  the minijet flux averaged over different flares \textit{is dominated by rare bright flares}. Another important statistical representation of the light curve is the power spectrum, $P(\nu)$, which measures its variability on different timescales: we predict $P(\nu)\propto \nu^{-2}$ for sufficiently high frequencies. 

The contradiction between low and high magnetization models described above calls into question the dominant mechanism of particle acceleration  in PWNe via  the stochastic shock acceleration and lends support to the alternative model, acceleration at relativistic reconnection events. If confirmed, this would be {\it a  major paradigm change in high energy astrophysics as a whole}, since PWNe, and especially the Crab PWN, is used as  prototypical source for studies of the high energy processes in the fields of Active Galactic Nuclei and Gamma Ray Bursts.

Reconnection physics is highly uncertain: it depends crucially on the
kinetic and geometric 
 properties of the plasma, which is very hard to test
observationally. 
Often,
various instabilities (based on 
inertial or cyclotron effects, 
 on ions or electrons) often seem to account  equally well
(with astrophysical accuracy) for the observed phenomena. 
 This situation may be contrasted with the shock acceleration schemes,
where  a qualitatively correct result for the spectrum of accelerated particles
can be obtained from simple {\it macroscopic } considerations, the shock jump conditions \citep[\eg][]{BlandfordEichler}.

\begin{figure}
\centering
\includegraphics[width=.99\linewidth]{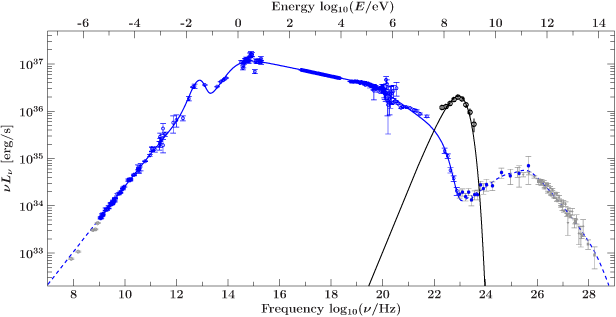}
\caption{Overplotted are single electron SED and the \Fermi\ flare data (black line and data) over data from a typical quiescent nebula spectrum and its model fit from \cite{2010A&A...523A...2M}.}
\label{X-gamma-flare} 
\vskip -.2truein
\end{figure}

 We  investigate the model in which 
high-energy particles are accelerated  at localized reconnection events, \eg\ on the equator on near the pole  \citep{2010MNRAS.405.1809L}.  Previously,   \cite{2012MNRAS.426.1374C} showed that a statistical model of reconnection mini-jets is  able to reproduce the  observed intensity fluctuations. \cite{2012MNRAS.426.1374C} argued that all of the high energy emission from the Cab nebula, both the nearly constant overall back ground and rare bright flares may be due to random stochastic  reconnection events. They relied on the observation by \cite{LyutikovUzdensky} that in case of relativistic reconnection the outflow velocity may be relativistic - even mildly relativistic outflows, \eg with the Doppler factor of only a few, will produce high brightness variations due to the high-power dependence of the observed luminosity on the bulk motion. 
For a mono-energetic spectrum, the flares are easily missed at lower energies, Fig. 
\ref{X-gamma-flare}.

\begin{figure}
%\begin{figure}[h!]
\centering
$
\begin{array}{cc}
\includegraphics[width=.45\linewidth]{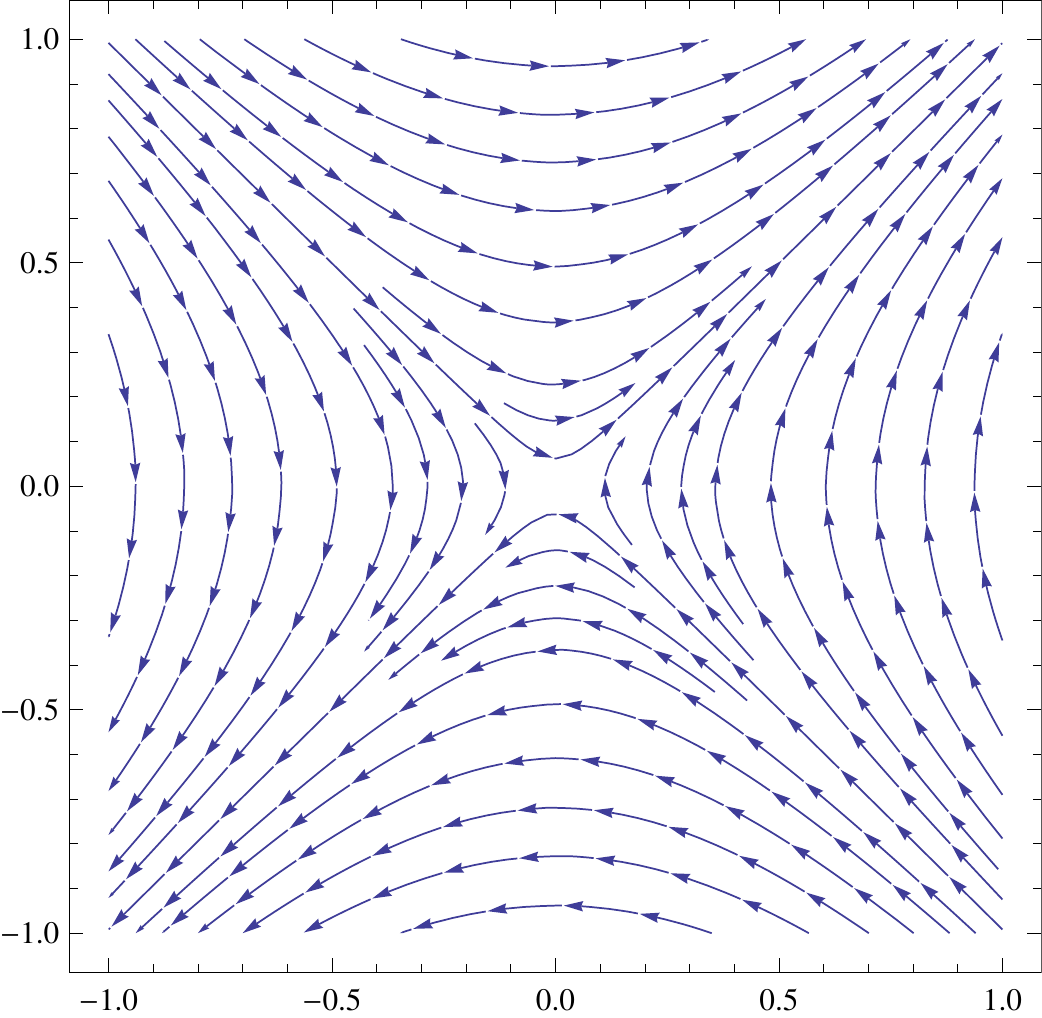} &
\includegraphics[width=.45\linewidth]{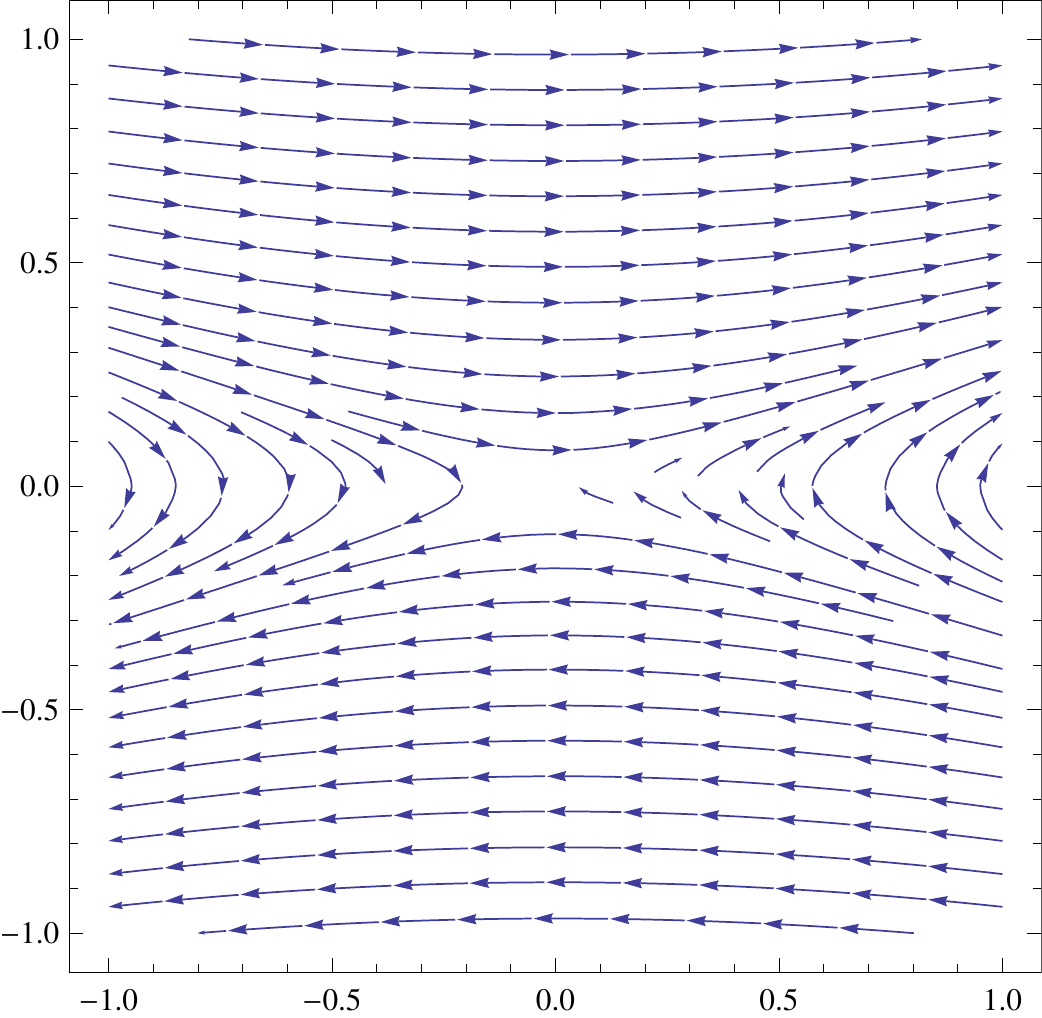} \\
\includegraphics[width=.22\textwidth]{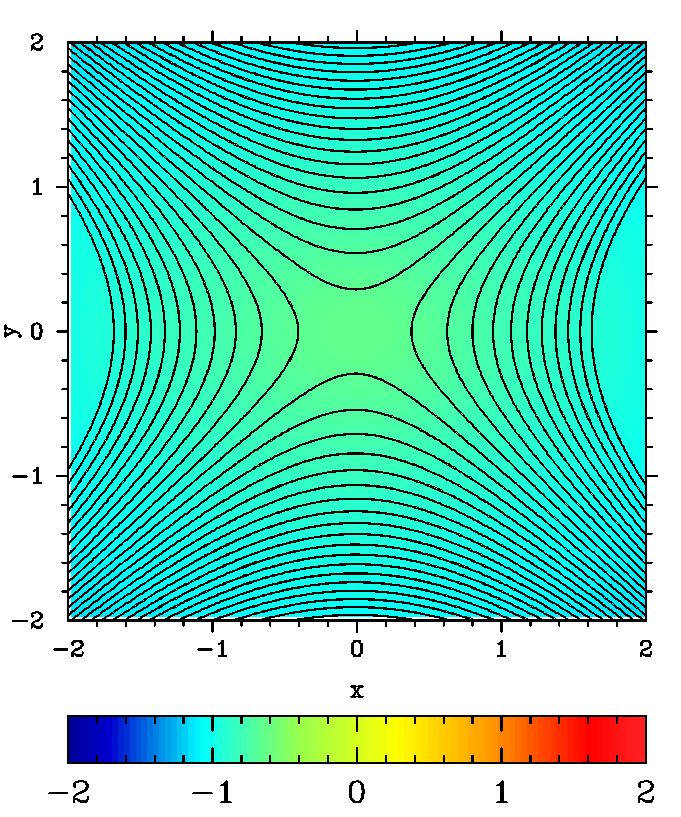}&
\includegraphics[width=.22\textwidth]{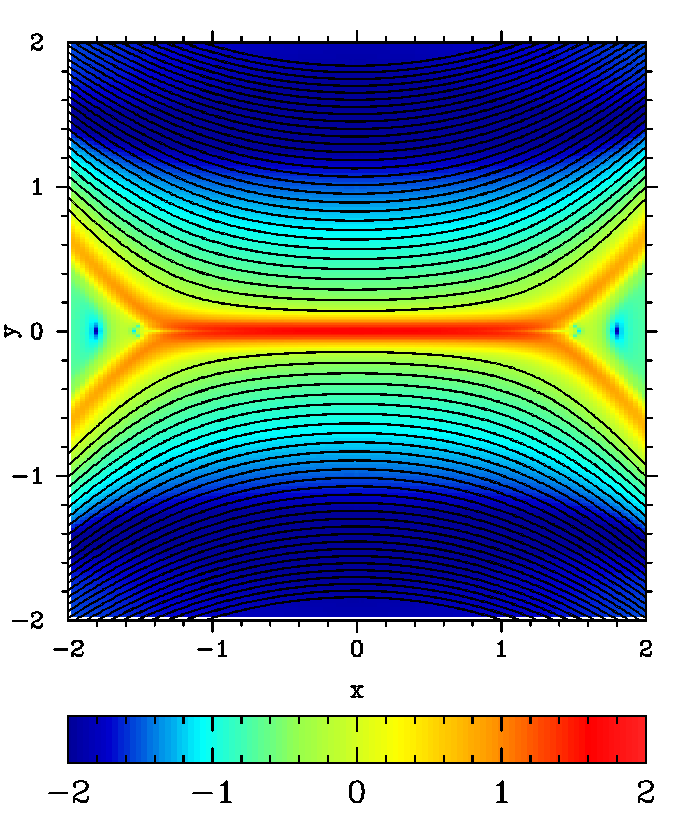}
\end{array}
$
\caption{{\it Upper row}. Analytical structure of the \Bf\ in the $x-y$ plane during X-point collapse in force-free plasma. The initial configuration on the left is slightly ``squeezed''. On dynamical time scale the X-point collapses to form  a current sheet, right figure. {\it Lower row}. Numerical simulations of the X-point collapse (Komissarov, priv. comm.).}
\label{B-field-coll} 
\end{figure}

The microphysics of the acceleration in relativistic reconnection layers has been 
previously addressed by \cite{2012ApJ...746..148C}, who considered a   {\it stationary} relativistic  reconnection regime, following qualitatively the stationary  Sweet-Parker model. But within the framework of stationary models of reconnection \citep[\eg][]{LyutikovUzdensky,2005MNRAS.358..113L} it is hard to achieve  relativistic inflow velocities \citep{2007ApJ...670..702Z}.
We are developing a model of explosive collapse in a highly magnetized plasma, Fig. \ref{B-field-coll}. The model has all the ingredients needed for Crab flares: explosive
dynamics on \Alfven (light travel) time scale (starting from smooth initial condition), development of high \Ef\ regions, with \Efs\  of the order of   \Bfs\   in the bulk and collimation of particles towards the low \Bf\ regions. 

\vskip -.5 truein

 \bibliographystyle{apj}
  \bibliography{/Users/maxim/Home/Research/BibTex}

\end{document}